\documentclass{article}
\usepackage[english]{babel}
\usepackage{amsmath,amssymb}
\usepackage[linktocpage=true]{hyperref}

\newcommand{\cM}{{ M}}

\begin{document}

   \vspace{1.8truecm}

 \centerline{\LARGE \bf {\sc Compactifying the Sen Action:}}
 \vskip12pt
\centerline{\LARGE \bf {\sc  Six Dimensions} }
 
 \vskip12pt


\vspace{1cm}

\vspace{1.0truecm}

 
\thispagestyle{empty}

\vspace{0cm}
  \centerline{
   {\large  {\sc Neil Lambert }\footnote{E-mail address: \href{mailto:neil.lambert@kcl.ac.uk}{\tt neil.lambert@kcl.ac.uk}}   and  {\sc Yuchen Zhou }}\footnote{E-mail address: \href{yuchen.1.zhou@kcl.ac.uk}{\tt yuchen.1.zhou@kcl.ac.uk}}       }

\vspace{1cm}
\centerline{${}^b${\it Department of Mathematics}}
\centerline{{\it King's College London }} 
\centerline{{\it London, WC2R 2LS, UK}}

\vspace{12pt}

\centerline{\sc Abstract}
\vspace{0.4truecm}
\begin{center}
\begin{minipage}[c]{360pt}{
    \noindent  The Sen action for self-dual fields has been generalised by Hull to include two metrics which allows it to be defined on generic manifolds. In this paper we consider Kaluza-Klein compactifications of this action. The existence of two metrics presents novel challenges as there are two   Kaluza-Klein towers of   fields. We show that   to    find  a consistent truncation  one must include   zero-modes associated to each of the two towers.  Although this naively leads to a doubling of the massless degrees of freedom we show that on-shell this is not the case. We also discuss a deformation of the Sen action to include an additional  form-field but which does not lead to new degrees of freedom on-shell but which arises naturally upon compactification.}
\end{minipage}
\end{center}
 
 \vskip2cm
 \begin{centerline}
 	{\it Dedicated to the memory of Kelly Stelle}
 \end{centerline}

 \newpage

\section{Introduction}

In $4k+2$ dimensions the representation theory of the Lorentz groups allows for self-dual fields, namely a field $2k$-form $A$ which satisfies $dA = \star dA$. There are many notable applications of these fields such as a chiral Boson in two-dimensions, the 2-form arising on an M5-brane in six-dimensions and 5-form Ramond-Ramond field of type IIB supergravity in ten-dimensions.  However constructing Lagrangians for these theories, and their subsequent quantization, is remarkably rich and subtle. A variety of approaches have been developed over almost the past half a century, for example see \cite{Siegel:1983es,Floreanini:1987as,Henneaux:1988gg,Pasti:1996vs,Belov:2006jd,Mkrtchyan:2019opf,Townsend:2019koy}.

A relatively recent approach was initiated by Sen in \cite{Sen:2015nph,Sen:2019qit} based on String Field Theory. This action in fact describes two self-dual fields. It's novelty is two-fold: The two self-dual fields have opposite sign kinetic terms but also while one couples to the physical spacetime metric $g$ the other couples to the Minkowskian metric $\eta$. The action is Lorentz invariant and quadratic in the fields and as such is amenable to quantization by path integral techniques \cite{Andriolo:2021gen, Lambert:2023qgs,Hull:2025rxy}.

The Sen action was generalised by Hull to include two independent metrics $g$ and $\bar g$  \cite{Hull:2023dgp}. Crucially his allows  the original Sen action to be defined on generic manifolds. It also opens up an interesting new arena of theories based on two independent, but non-dynamical, metrics. For example it admits two different diffeomorphism-like symmetries; one associated to each metric. Traditional diffeomorphisms arising from     coordinate transformations arise as the diagonal subgroup. The appearance of two metrics also  allows one to disentangle the dynamics of one form field from the other.    In particular there are two independent conserved  energy-momentum tensors \cite{Hull:2025bqo}. The problem of the wrong sign kinetic term can also be alleviated by choosing $\bar g$ to have the opposite signature to $g$.
 
 Thus the Sen action is novel in several ways. In this paper we will explore  its compactification on a generic manifold. This too  is rather subtle and requires new considerations in an  otherwise old topic. We note that compactifications of the Sen action have been considered before \cite{Sen:2019qit,Andriolo:2020ykk,Aggarwal:2025fiq} but only on tori in a particular coordinate system where the unphysical metric $\bar g$ is constant.  Here we wish to address issues that arise where the compact manifold has two non-trivial metrics.

 The main issue we will encounter is that there are two distinct Kaluza-Klein (KK) towers of fields corresponding arising from the two Laplacians of $g$ and $\bar g$. In traditional KK constructions one has a single KK tower and restricts  to its  zero-modes. We will see that this simple ansatz is not valid and in particular does not provide a consistent truncation, in the sense that   no solutions to equations of motion of the lower-dimensional theory,   lift to solutions of the original theory. We will see that these problems can be overcome by using a novel KK anstaz that invokes zero-modes of both Laplacians. Alternatively, from the point of view of an expansion in a single KK tower, we keep the zero modes along with particular one-dimensional towers of non-zero KK modes that correspond to the zero-modes of the second KK tower.

One might   attempt to resolve this problem by simply setting the two internal metrics equal, leaving just one KK tower. However it is unsatisfactory to have a theory constructed with two independent metrics which must then be related if we wish to compactify it. In addition, in a quantum treatment of the Sen action, the existence of two separate metrics is a crucial feature that allows the path-integral to be well-defined \cite{Andriolo:2020ykk,Hull:2025rxy}. Therefore it is desirable to explore the full theory and its structure and then examine its consequences. We also note that the Sen action  has its origins in String Field Theory where it was shown that  the resulting dynamics is independent of the choice of background fields, including the metrics $\bar g$ and $g$ \cite{Sen:2017szq,Hull:2025mtb}. In particular different  choices of $\bar g$ or $g$ are related by a field redefinition, provided that that the string fields are on-shell, including the metric equations of motion. However in the applications explored here we are not working within the context of String Field Theory or even supergravity and both metrics are taken to be fixed and non-dynamical. 
 
Our analysis also reveals a generalization of the Sen action to include an additional $2k$-form field. In particular even if this field is not present in the original theory it will appear upon compactification,  arising from the additional KK modes associated to the second metric.  While one might expect this field to introduce new degrees of freedom, it doesn't, as it can be removed on-shell by a field redefinition. However it does not seem possible to make this redefinition at the level of the action. Thus our results are reminiscent of the expectation from String Field Theory in the sense that the dynamical modes arising from a second KK tower do not lead to additional on-shell modes.

   We note that consistent KK truncations   have a long history. For a recent discussion of consistent truncations  in the context of supergravity see \cite{Lin:2024eqq,Lin:2025ucv,Lin:2026rfc}.
 In this paper we will take a consistent truncation to mean that a solution of the lower dimensional equations leads, via the compactificion ansatz, to solution of the higher dimensional equations of motion. However, as we will see here, there is a subtlety: while it may be the case that the compactification ansatz leads to a set of lower dimensional equations of motion which are consistent, in the sense just described, one can also substitute the compactification ansatz directly into the higher dimensional action and then integrate over the internal manifold. As we will see here  the  dimensionally reduced action leads to a more general set of lower-dimensional equations of motion. While the equations of motion arising from the dimensional reduced  action include  consistent solutions they also  allow for inconsistent solutions.  Similar behaviour has been noticed before in the context of sphere reductions of supergravity \cite{Lu:2006dh}.

 The rest of this paper is organised as follows. In section \ref{sec: Sen}  we review the Sen action and its extension to two metrics due to Hull. We will also introduce   the generalization to include a second $2k$-form field which turns out to carry no new on-shell degrees of freedom. In section \ref{sec: 6D} we will start our analysis of a KK reduction explaining in details the issues that arise. For simplicity we first consider  the case of reduction of the six-dimensional theory on ${\mathbb CP}^2$. Here we describe in  detail the issues that are faced by having two metrics and their resolution. We  then apply this analysis to the case of the M5-brane on $K3$. In section \ref{sec: RS} we consider compactification of a six-dimensional self-dual theory on a Riemann surface. Section \ref{sec: Conclusion} contains our conclusions. We will address the reduction of type IIB supergravity in future work.

\section{The Action}\label{sec: Sen}

The Sen action \cite{Sen:2015nph,Sen:2019qit} is defined in $4k+2$ dimensions.   In the presence of a source $J$ and coupled to a general background metric \cite{Hull:2023dgp} it is  
\begin{align}
S = - \int \Big(&\frac12 dB\wedge \bar\star  dB + 2 H\wedge dB-(H+J)\wedge \cM(H+J)\nonumber\\
&+2H\wedge J - \frac12 J\wedge \bar\star J\Big)	\ .
\end{align}
Here $B$ is a $2$-form, $H$ a $(2k+1)$-form which is self-dual with respect to a Lorentzian-signature metric $\bar g$; $H=\bar\star H$.  Furthermore $\cM$ is linear map constructed to take $\bar \star$-self-dual forms to $\bar \star$-anti-self-dual forms and in particular satisfies:
\begin{align}
\cM(H) = \frac12(1-\bar\star)	\cM\left(\frac12(1+\bar\star)H\right)\ .
\end{align}
Its role is to ensure that
\begin{align}
H + \cM(H) = \star \left(H + \cM(H)\right)	\ ,
\end{align}
where $\star $ is the Hodge dual associated to a second, physical  metric $g$ that is  independent of $\bar g$. For a general construction of $\cM$ see \cite{Sen:2019qit,Andriolo:2020ykk,Hull:2023dgp}.

The equations of motion of this action take the form
\begin{align}
0 &=   d\left(-\bar\star dB+2H\right)  \label{eqofm1}\\
0 &= dB-\bar \star dB - 	2\cM(H+J) +J-\bar \star J \label{eqofm2}\ .
\end{align}
Defining 
\begin{align}
F &= H+\tfrac12(1+\bar\star )J+\cM(H+J)  \\ 
G &= H-  \frac12(1+\bar\star )dB \ , 
\end{align}
the equations of motion can be rewritten as
\begin{align}\label{eqM}
dG & = 0 \\
dF & =  dJ \ .
\end{align}
In this way we find two dynamical self-dual forms with respect to the each of the two metrics; $F= \star F$  and $G = \bar \star G$.   

We can include an independent source $\bar J$ for $G$   by taking
 \begin{align}
S = - \int \Big(&\frac12 dB\wedge \bar\star  dB + 2 H\wedge dB-(H+J-\bar J)\wedge \cM(H+J-\bar J)\nonumber\\
&+2dB\wedge\bar J+2H\wedge (J-\bar J) - \frac12 (J-\bar J)\wedge \bar\star (J-\bar J)\Big)	\ .
\end{align}
This leads to 
\begin{align}\label{eqofm5}
 0&=   d\left(-\bar\star dB+2H-2\bar J\right)  \\
0&= dB-\bar \star dB - 	2\cM(H+J-\bar J) +(1-\bar \star )(J-\bar J) \label{eqofm6}  \ .
\end{align}
and hence
\begin{align}\label{eqM}
dG & = d\bar J \\
dF & =  dJ \ .
\end{align}
 where now
\begin{align}
F &= H+\tfrac12(1+\bar\star )(J-\bar J)+\cM(H+J-\bar J) \\ 
G &= H-  \frac12(1+\bar\star )dB \ , 
\end{align}

Note that we could also consider a Sen action with $\bar \star H=-H$ and $\cM$ defined as mapping $\bar g$-anti-self-dual forms to $\bar g$-self-dual forms. This leads to an analogous theory but where the forms $F$ and $G$ are now anti-self-dual with respect to $ g$ and $\bar g$ respectively. One could also consider non-interacting copies of Sen actions with different choices of sign of self-duality constraints. These will all play a role in what follows.   
 
\subsection{A Generalised   Action}\label{sect: GSen}

Before looking at KK compactification  we first want to 
note    an interesting extension of the Sen action. Let us introduce  an additional  $2k$ form $A$ and extend the action to
\begin{align}\label{GSen}
S' = - \int \Big(&\frac12 dB\wedge \bar\star  dB + 2 H\wedge dB-H\wedge \cM(H)  \Big)\nonumber \\
& -\frac12\lambda\int  \Big(dA - H-\cM(H)\Big)\wedge\bar \star \Big(dA - H-\cM(H)\Big)	\ ,
\end{align}
where   $\lambda$ an arbitrary constant.  The equations of motion are
\begin{align}
0&= d\left( H-\frac12\bar\star dB \right)  \\
0&=  d\bar\star(H+\cM(H) -dA) \\
0&= \left(2 \cM(H)-d B+\bar\star dB  \right) - \frac12 \lambda\left( d  A-\bar\star dA+{\cM}\left(d  A+\bar\star dA\right)-4\cM(H)\right) \ . 
\end{align}
Naively one might expect that by including $A$  we have introduce   new degrees of freedom but this is not the case. To see this we note that from the second equation 
 we can  write, at least locally, 
\begin{align}
dA &= H+\cM(H)	+\bar\star d\bar A\ ,\qquad 
\bar\star dA = H-\cM(H)	+  d\bar A\ ,
\end{align}
for any  $2k$-form $\bar A$. The remaining equations become
\begin{align}
0&= d\left( H-\frac12\bar\star dB - \frac12 dB \right)  \\
	0&= \left(2 \cM(H)-d B+\bar\star dB  \right) -  \frac12\lambda\left( \bar\star d\bar A- d\bar A+{\cM}\left(d  \bar A+\bar\star d\bar A\right) \right)  \ ,
\end{align}
which can be written as 
\begin{align}
0&= d\left( H'-\frac12\bar\star dB' - \frac12 dB' \right)  \\
	0&=  2 \cM(H')-d B'+\bar\star dB'   \ ,
\end{align}
where 
\begin{align}
	H' & = H -\frac14\lambda d\bar A-\frac14\lambda\bar\star d\bar A\\
	B' & = B -  \frac12\lambda \bar A \ .
\end{align}
In this way we recover the same set of solutions 
\begin{align}
dG'&=0  \\
dF'&=0	\ ,
\end{align}
with $G'=\bar\star G'$, $F'=\star F'$ and 
\begin{align}
G'  &= H' - \frac12 dB'-\frac12 \bar \star dB	'\\
F' & = H' + \cM(H') \ .
\end{align}
Thus on-shell the local degrees of freedom carried by $A$ can be absorbed into a field redefinition of $H$ and $B$. However such a field redefinition doesn't seem possible at the level of the action.


\section{Compactification  }\label{sec: 6D}

To start our discussion of compactification we want to consider the Sen action on a  manifold of the form $\Sigma\times {\cal K}$. We assume product metrics:
\begin{align}
\bar g  &= (\bar \gamma,\bar \kappa) \\
  g  &= (  \gamma,  \kappa)\ ,
\end{align} 
We denote the Hodge stars as $\bar\star$ and $\star$ respectively and it should be understood from the context whether  they act on forms on $\Sigma$,  ${\cal K}$ or $\Sigma\times {\cal K}$. In what follows we always take $\cal K$ to be the compactification manifold and we look for the low energy theory on $\Sigma$. 
 
 In order to be concrete it is helpful to consider a specific example. Arguably 
   the simplest case consists of taking   the six-dimensional action reduced to two-dimensions on  ${\cal K}={\mathbb CP}^2$. Let us discuss this  first to illustrate the novelties we encounter. 
 
In KK reductions we expand the fields in a basis of   2-forms on ${\mathbb CP}^2$. Looking at the $B$-field kinetic terms it is natural to consider the expansion
\begin{align}
	B &= b \bar\omega + \sum_n b_n\bar \phi_n\ .
\end{align}
Here $\bar \omega,\bar\phi_n$ are a basis of two-forms that are eigenvalues of the Laplacian obtained from $\bar g$:
\begin{align}
\bar\star d\bar\star d\bar \phi_n +d\bar\star d\bar\star \bar \phi_n = -m^2_n\bar\phi_n	 \ .
\end{align}
We have separated out the zero-mode $\bar\omega$, which is unique on ${\mathbb CP}^2$, and can be taken to satisfy 
\begin{align}
	d\bar\omega=d\bar\star\bar \omega=0\ ,
\end{align}
and also
 \begin{align}
 \bar\omega=\bar\star\bar \omega \ .	
 \end{align}
As is well known eigenmodes with different eigenvalues are orthogonal:
\begin{align}
\int_{{\mathbb CP}^2} \bar \omega\wedge\bar\star\bar\phi_m =\int_{{\mathbb CP}^2} \bar \phi_n\wedge\bar\star\bar\phi_m =0	\ ,
\end{align}
for $m\ne n$. 
Looking only  at the kinetic terms for $B$ we find
\begin{align}
 \frac12 \int_{\Sigma\times {\mathbb CP}^2}   dB\wedge \bar \star dB =  \frac12 \int_{\Sigma }db \wedge \bar\star db+ \frac12 \sum_n
	\int_{\Sigma } \left( db_n \wedge \bar\star db_n + m_n^2b_n \wedge \bar\star  b_n\right)\ .
\end{align}
So far this is just an expansion of the higher dimensional field in a particular basis and nothing has been lost. The key idea behind KK theory is that the non-zero modes become massive fields and hence at low energy, below the scale set by the lowest non-zero mode, we can neglect the higher modes $b_n$. Thus we truncate the action to 
\begin{align}
 \frac12 \int_{\Sigma\times {\mathbb CP}^2}   dB\wedge \bar \star dB \to   \frac12 \int_{\Sigma }db \wedge \bar\star db \ .
 \end{align}
 Next we could make a related expansion for $H$:
 \begin{align}
	  H &= h\wedge \bar\omega+ \sum_n h_n\wedge \bar \phi_n\ ,
\end{align}
  where $h $ and $h_n$ are 1-forms on $\Sigma$. 
  The $H\wedge dB$ term in the action will then also nicely reduce to diagonal form which doesn't mix different KK levels  and we can simply truncate to the zero-mode $H=h\wedge \bar\omega$.  
  
  However an important question is whether or not this truncation to the massless fields is consistent with the higher-dimensional theory. A consistent truncation is one in which the zero-modes decouple from the non-zero modes. In particular in a consistent trucation a solution to the lower-dimensional equations of motion, leads, via the KK ansatz, to a solution of the original higher dimensional ones. Said differently it is possible that the equations of motion of the higher dimensional theory do not allow one to simply set the non-zero modes to zero. 
  
 In our case problems arise  when we consider $\cM(H)$. This is required to satisfy 
  \begin{align}
  h\wedge \bar\omega + {\cM }(h\wedge \bar\omega) = \star \left( h\wedge \bar\omega + {\cM }(h\wedge \bar\omega) \right)	\ ,
  \end{align}
but there is no reason why ${\cM }(h\wedge \bar\omega)$ should be restricted to  zero-modes of the $\bar g$-Laplacian. Indeed we will show this below. Hence we can't truncate to zero-modes alone.

 Since we have two metrics we also have a second KK tower:
 \begin{align}
 \star d \star d  \phi_n +d \star d\star \phi_n = -m^2_n \phi_n	 \ ,
\end{align}
 with a zero-mode   $\omega=\star \omega $ such that
 \begin{align}
 d \omega = d\star \omega =0	\ .
 \end{align}  
Therefore we can also consider    an alternative  KK expansion of the form 
 \begin{align}
	B &=   b  \omega + \sum_n   b_n  \phi_n\ .
\end{align}
However we now find the kinetic term will no longer be diagonal, {\it i.e.} it will mix fields with different mode numbers meaning that it is inconsistent to truncate to the zero-modes.

For guidance on how to proceed let us postpone our discussion of the action and instead  look at the equations of motion (\ref{eqM}). We are looking for  an ansatz for the KK zero-modes such that 
\begin{align}\label{wish}
G& = G_0\wedge \bar \omega\\
F & = F_0\wedge \omega	\ ,
\end{align}
with $ G_0=\bar\star G_0$ and $F_0 = \star F_0$. In this case the equations $dG=dF=0$ reduce to simply $dG_0=dF_0=0$  (in the absence of sources). This would provide a consistent reduction: meaning solutions to the two-dimensional equations $dF_0=dG_0=0$ lift to solutions to the   six-dimensional equations $dF=dG=0$.  

It is critical in the Sen system that  $G$ decouples from the metric $g$ and hence one might think that we do not want to expand $B$ and $H$ in terms of harmonic forms of $g$ (however we will see that we will have to loosen this condition). Therefore to start  we consider a  KK expansion based on  eigenmodes of the $\bar g$-Laplacian and restrict to the zero modes: \begin{align}
	B &= b \bar\omega  \\
	H &= h\wedge \bar\omega \ .
\end{align}
The condition $H=\bar\star H$ implies $\bar\star h=h$.
 In this way we find
 	\begin{align}
 	G &= \left(h  - \tfrac12(1+\bar\star)db\right)\wedge\bar\omega	\\
 	F & = h\wedge \bar\omega + \cM(h\wedge \bar\omega)\ .
 	\end{align}
Thus we have 
$G_0  =h  - \tfrac12(1+\bar\star)db $ 
and 
the equation $dG=0$ simply reduces to $d(h- \tfrac12(1+\bar\star)db)=0$. Furthermore $G_0$ is independent of the metric $g$ as desired.

On the other hand for $F$ we require that 
\begin{align}\label{F1}
	h\wedge \bar\omega + \cM(h\wedge \bar\omega)
 = F_0\wedge \omega \ ,
\end{align}
for some $F_0$ with $\star F_0=F_0$. However this is not possible. To see this we expand (recall that $(\bar \omega,\bar \phi_n)$ are a basis of 2-forms on ${\mathbb CP}^2$)
\begin{align}
\cM(h\wedge \bar\omega)
= m(h)\wedge  \bar  \omega +\sum_n m_n(h)\wedge  \bar \phi_n	\ .
\end{align}
Next 
take the wedge product of (\ref{F1}) with $ \bar \omega$ and integrate over ${\mathbb CP}^2$:
\begin{align}
(h +m(h)) \int_{ {\mathbb CP}^2} 	 \bar\omega\wedge\bar\omega  = F_0 \int_{ {\mathbb CP}^2} \omega\wedge \bar \omega \ ,
\end{align}
where we have used the fact that a self-dual 2-form wedge an anti-self-dual 2-form vanishes. Next we observe that   $\omega$ and $\bar \omega $   both represent the same cohomology class and hence there exists a 1-form $\chi$ such that 
\begin{align}
\omega = \bar\omega + d\chi	\ .
\end{align}
Therefore 
\begin{align}
\int_{ {\mathbb CP}^2} \omega\wedge \bar \omega  = \int_{ {\mathbb CP}^2} \bar \omega\wedge \bar \omega \ ,
\end{align}
and it follows that $F_0=h+m(h)$. Substituting back into (\ref{F1}) we find
\begin{align}
\cM(h\wedge \bar\omega)
 = h\wedge (\omega-\bar\omega)	+m(h)\wedge \omega\ ,
\end{align}
but this is not $\bar g$-anti-self-dual. In particular
\begin{align}
(1+\bar\star)\cM(h\wedge \bar\omega) = 	
h\wedge (\omega+\bar\star \omega - 2 \bar\omega)	+m(h)\wedge (\omega-\bar\star\omega)\ ,
\end{align}
and we note that $h$ is $ \bar\star $-self dual and $m(h)$ is $ \bar\star $-anti-self dual. Therefore this can only vanish if $ \bar \omega=\tfrac12(\omega+ \bar\star \omega)$  and $\omega=\bar\star\omega$ which implies $\omega=\bar\omega$.

To continue we must revisit our KK ansatz. Let us instead determine $H$ and $\cM(H)$ by the demanding that we reduce to a two-dimensional Sen-system:
\begin{align}
H + \cM(H) = (h + m(h))\wedge \omega	\ ,
\end{align}
with  $\bar\star  h=h$ and $\bar\star m(  h)=-m(  h)$ chosen so that  $h + m(h)=\star (h + m(h))$.  
We can find $H$ by projecting on the the $\bar g$-self-dual part:
\begin{align}
H & = \frac12 (1+\bar\star )	\left((h + m(h))\wedge \omega\right)\nonumber	\\
& = h \wedge \frac12 (1+\bar\star )	\omega + m(h)\wedge \frac12 (1-\bar\star )	\omega\nonumber \\
&= h\wedge\bar\omega	+\tfrac12(1+\bar\star)\left((h+m(h) )\wedge (\omega-\bar\omega)\right)\ .
\end{align}
Thus we see that we need to include an additional term in $H$ involving the $g$-self-dual form $\omega$. Similarly we can determine $\cM(H)$ by projecting onto the $\bar g$-anti-self-dual part:
\begin{align}
	\cM(H)& =  \frac12 (1-\bar\star )	(h + m(h))\wedge \omega	\nonumber \\
	&=  \frac12m(h)\wedge (1+\bar\star)\omega+  \frac12h\wedge (1-\bar\star)\omega\ .
	\end{align}
	
Thus we have found an ansatz for $H$ that leads to the desired form of $F$. However the price we pay is that now
\begin{align}
G =\tfrac12(1+\bar\star)\left((h+m(h) )\wedge (\omega-\bar\omega) \right)+\left(h - \tfrac12(1+\bar\star)db\right)\wedge\bar  \omega	\ .
\end{align}
Although this is $\bar\star$-self-dual it depends  on $g$, which goes against the spirit of the Sen action. More problematically the equation of motion $dG=0$ isn't solved by the two-dimensional equation $d\left(h - \tfrac12(1+\bar\star)db\right)=0$. Rather only the averaged equation of motion is solved:
\begin{align}
\int_{{\mathbb CP}^2}dG\wedge\bar\omega=0	\ .
\end{align}
So this is does not provide a  consistent truncation.

To obtain a consistent truncation we can also extend in our ansatz for $B$:
\begin{align}
B & = b   \bar \omega  +  a(\omega-\bar\omega)\ ,
\end{align}
to include an additional two-dimensional scalar field $a$. 
This leads to 
\begin{align}
G&= \tfrac12(1+\bar\star)\left((h+m(h) -d  a)\wedge (\omega-\bar\omega) \right)+\left(h - \tfrac12(1+\bar\star)db\right)\wedge\bar  \omega\\
F& =  	\left(h+m(h)\right)\wedge \omega\ .
\end{align}
Let us now look at the equations of motion. In particular (\ref{eqofm2}) becomes
\begin{align}
0= d(b-a)\wedge \bar\omega   &- \bar\star d(b-a)\wedge \bar\omega    + da\wedge \omega-\bar\star d a\wedge\bar\star\omega\nonumber\\
&  -m(h)\wedge (1+\bar\star)\omega -h\wedge (1-\bar\star)\omega\ .
\end{align}
For general choices of independent metrics the terms with $\bar\omega$, $\omega$ and $\bar\star \omega$ must all vanish separately leading to\begin{align}
d(b-a)    &= \bar\star d(b-a)\ , \label{eqofm3} \\da &= h+m(h)\ ,\\
\bar\star da & = h-m(h) .
\end{align}
Happily the  second and third equations agree  and these imply $dF=0$ as required. 
Next we evaluate (\ref{eqofm1}) which is equivalent to $dG=0$ but now $G$ has simplified to 
\begin{align}
G = 	\left(h - \tfrac12(1+\bar\star)db\right)\wedge\bar  \omega\ ,
\end{align}
and hence $dG=0$ becomes simply
\begin{align}
d	\left(h - \tfrac12(1+\bar\star)db\right)=0 ,
\end{align}
as expected.

To summarize we have found a  KK reduction ansatz that admits a consistent reduction:
\begin{align}
	B & = b  \bar \omega  +   a (\omega-\bar\omega)\ ,\\
	H & = h\wedge\bar\omega	+\tfrac12(1+\bar\star)\left((h+m(h) )\wedge (\omega-\bar\omega)\right)\ ,\\
\cM(H) & =  \frac12m(h)\wedge (1+\bar\star)\omega+  \frac12h\wedge (1-\bar\star)\omega\ ,
\end{align}
where  $h =\bar\star h$, $m(h)=-\bar \star m(h)$ is constructed so that $(h +m(h))=\star(h +m(h))$.

  We note that   the two form $\omega-\bar\omega =d\chi$ is only made up of non-zero modes of the $\bar \kappa$-Laplacian:
\begin{align}
\int_{{\mathbb CP}^2} d\chi\wedge \bar\star   \bar\omega  = 0\\
\int_{{\mathbb CP}^2} d\chi\wedge \bar\star \bar \phi_n    \ne 0 \ .
\end{align}
Thus,  in addition to the truncation to the familiar zero-mode,  we have extended the naive KK ansatz by including a one-dimensional tower of non-zero modes in an expansion based on the eigen-forms of the $\bar g$-Laplacian. In particular the additional non-zero mode  is precisely the zero-mode of the $g$-Laplacian. 

It is also interesting to note that to solve (\ref{eqofm3}) we can write $b = a + c$ with $dc=\bar\star dc$ so that the expression for $G$ now becomes
\begin{align}
G = 	-\bar\star dc \wedge\bar  \omega ,
\end{align}
and this automatically satisfies $dG=0$ as $d\bar\star dc = d^2c=0$. 
A natural  special case is $c=0$  leading to $G=0$ and also  $B=b\omega$, {\it i.e.} we expand $B$ entirely in terms of the zero mode of the $g$-Laplacian (although $H$ is expanded in both harmonic forms).   
 
Next let us   implement this KK reduction at the level of the action. We find
\begin{align}
S = &-\bar V\int_\Sigma \frac12 db\wedge \bar\star db + 2h\wedge db  - h\wedge  m(h)	\nonumber\\
 &- \Omega \int_\Sigma \frac12 d  a\wedge \bar\star d  a   + \left(h-m(h)\right)\wedge   d  a -h\wedge  m(h)	\ ,
\end{align}
where 
\begin{align}
\bar V = \int_{{\mathbb CP}^2}	\bar\omega\wedge\bar \star \bar\omega
\qquad \Omega = \int_{{\mathbb CP}^2}	(\omega-\bar \omega)\wedge\bar \star (\omega-\bar \omega)\ .
\end{align}
This   action can be re-written using as:
\begin{align}
S = &-\bar V\int_\Sigma \frac12 \left(db-2h\right)\wedge \bar\star  \left(db-2h\right)  - h\wedge  m(h)	\nonumber\\
 &- \Omega \int_\Sigma \frac12 \left(d  a -h-m(h)\right)\wedge \bar\star  \left(d  a -h-m(h)\right)\ .
 \end{align}
Thus we land on the generalised Sen action that we discussed in section \ref{sect: GSen} above. As noted there, the additional field  $a$ does not lead to additional degrees of freedom  on-shell as it can be absorbed into $b$ and $h$:
\begin{align}
b' &= b- \frac{\Omega}{2\bar V}\bar a\\
h' & = h- \frac{\Omega}{4\bar V}(d  \bar a+\bar\star
 d\bar a)	\ .
\end{align}
Note that the general solution, where $\bar a\ne 0$, does not lift to a solution of the six-dimensional equations of motion. However it would if we adjusted the KK ansatz to involve $b'$, $a$,  and $h'$ instead of $b$, $a$ and $h$. 

Lastly we can consider what would happen if  we started with this generalised Sen action (\ref{GSen}) and reduce it with the   ansatz
\begin{align}
A = a_0\omega 	\ .
\end{align}
In this case we find the additional term
\begin{align}
S = &-\bar V\int_\Sigma \frac12 db\wedge \bar\star db + 2h\wedge db  - h\wedge  m(h)	\nonumber\\
 &- \Omega \int_\Sigma \frac12 d  a\wedge \bar\star d  a   + \left(h-m(h)\right)\wedge   d  a -h\wedge  m(h)		\nonumber\\
 &-  \Omega_0  \int_\Sigma \frac12 d  a_0\wedge \bar\star d  a_0   + \left(h-m(h)\right)\wedge   d  a_0 -h\wedge  m(h)	\ ,
\end{align}
where $\Omega_0 = \lambda(\bar V+\Omega)$ arises from
\begin{align}
\int_{{\mathbb CP}^2} \omega\wedge\bar\star \omega  = 	\int_{{\mathbb CP}^2} (\bar \omega+d\chi)\wedge\bar\star (\bar \omega+d\chi) = \bar V + \Omega \ .
\end{align}
In this way we find two additional modes $a_0$ and $a$ each with the same equation of motion. The equations of motion are all solved by taking $a=a_0$ with $da = h+m(h)$ in which case all terms involving $\Omega$ and $\Omega_0$ drop out and we recover the original Sen system. As above  the inclusion of more general solutions can be locally cancelled by an on-shell redefinition of $h$ and $b$. 

\subsection{Including Sources}

Let us now include the sources  $J$ and $\bar J$. Since these are expected to arise from other fields in the theory they will have the standard KK ansatz:
\begin{align}
\bar J = \bar j\wedge \bar\omega	\qquad J =  j\wedge  \omega	\ .
\end{align}
In order to absorb the factors of $J-\bar J$ in $F$ we modify the KK ansatz for $H$ to  
\begin{align} 	H & = h\wedge\bar\omega	+\tfrac12(1+\bar\star)\left((h+m(h) )\wedge (\omega-\bar\omega)\right) -\frac12(1+\bar\star)(j\wedge\omega -\bar j\wedge\bar\omega)\ .
\end{align}
In this way we find
\begin{align}
F = (h+m(h))\wedge \omega	\ ,
\end{align}
as before. On the other hand we now find
\begin{align}
G = 	&\left(h - \tfrac12(1+\bar\star)db-\tfrac12(1+\bar\star)(j-\bar j)\right)\wedge\bar  \omega\nonumber\\
&+\tfrac12(1+\bar\star)\left((h+m(h) -d  a -j)\wedge (\omega-\bar\omega)  \right) \end{align}
With this shift the equation (\ref{eqofm6}) becomes
\begin{align}
0= d(b-a)\wedge \bar\omega   &- \bar\star d(b-a)\wedge \bar\omega    + da\wedge \omega-\bar\star d a\wedge\bar\star\omega\nonumber\\
&  -m(h)\wedge (1+\bar\star)\omega -h\wedge (1-\bar\star)\omega + (1-\bar \star)(j\wedge \omega - \bar j\wedge \bar\omega)\ .
\end{align}
From here we read off that
\begin{align}
0&=d(b-a)	- \bar\star d(b-a) -(1-\bar\star)\bar j\label{eqofmA} \\
0&=da - h -m(h) +  j\label{eqofmB}\\
0 & = -\bar\star da + h -m(h) - \bar \star  j\label{eqofmC}\ .
\end{align}
The last two equations are equivalent and tell us that $d(h+m(h))= dj $, {\it i.e.} they imply $dF = dJ$, as required. 
Furthermore now $G$ reduces to
\begin{align}
G = 	\left(h - \tfrac12(1+\bar\star)db-\tfrac12(1+\bar\star)(j-\bar j)\right)\wedge\bar  \ ,\omega\end{align} 
and, using the above equations, one finds that (\ref{eqofm5}) is equivalent to
\begin{align}
dG &= d\left(h -\tfrac12(1+ \bar\star )db-\tfrac12(1+ \bar\star )(j-\bar j) \right)\wedge\bar\omega\nonumber \\
& = d \bar j  \wedge\bar\omega\nonumber\\
&= d\bar J \ ,\end{align}
where in the second line we have used (\ref{eqofmA}) and (\ref{eqofmB}). Thus we see that the sources can also be included in a consistent truncation. However in the rest of this paper, in the interests of clarity, we will not consider sources.

\section{The M5-brane on $K3$}

Let us now look at a more involved case: the M5-brane on $K3$. It was first proposed in \cite{Hull:1994ys, Witten:1995ex} that the M5-brane wrapped on $K3$ gives rise to the Heterotic String compactified on ${\mathbb T}^3$. The associated reduction at the level of the worldvolume action was shown in \cite{Cherkis:1997bx} using the PST formulation.\footnote{ In  \cite{Andriolo:2020ykk} the Sen formulation was also studied on $K3 $ but only for the field $F$ using  the Hamiltonian which avoided the fact that $\bar g=\eta$ is not defined on $\Sigma\times K3$.} Here we will  adapt our previous discussion of compactification on ${\cal K}={\bf CP}^2$ to ${\cal K}=K3$. 

In this case $K3$ has 22 harmonic 2-forms $\bar \omega_I$ with respect to $\bar g$ as well as a second set of 22 harmonic 2-forms $\omega_I$ with respect to $g$. These are closed and satisfy the self-duality relation
\begin{align}
\bar \star \bar \omega_I = \eta_{IJ}\bar \omega_J\qquad  	\star \omega_I = \eta_{IJ}\omega_J \ ,
\end{align}
where $\eta_{IJ}$ takes the form $\eta_{IJ} = {\rm diag}(-1,-1,-1,1,...,1)$, {\it i.e.} 3 are anti-self-dual and 19 are self-dual. These can be chosen such that
\begin{align}
\int_{K3} \bar \omega_I\wedge\bar\star\omega_J = \bar V\delta_{IJ}\ .	
\end{align}
Thus our KK ansatz is
\begin{align}
B &= b^I\bar\omega_I + a^I(\omega_I-\bar\omega_I)\ ,\\
H & = h^I\wedge\bar\omega_I	+\tfrac12(1+\bar\star)\left((h^I+m^{IJ}(h^J) )\wedge (\omega_I-\bar\omega_I)\right)\ ,\\
\cM(H) & =  \frac12m^{IJ}(h^J)\wedge (1+\bar\star)\omega_I+  \frac12h^I\wedge (1-\bar\star)\omega_I\ .
\end{align}
The condition $H=\bar\star H$ and $\cM(H)=-\star\bar \cM(H)$ now leads to the conditions
\begin{align}
\bar\star h^I &= \eta_{IJ} h^J \ ,\\
\bar\star m^{IJ}(h^J) &=	-\eta_{IJ}m^{JK}(h^K)\ .
\end{align}
And $m^{IJ}(h^J)$ is constructed so that
\begin{align}
h^I + 	m^{IJ}(h^J) = \eta_{IJ}\star \left(h^J + 	m^{JK}(h^K)\right)\ .
\end{align}
The six-dimensional equations of motion are
\begin{align}
0& = d\left(h^I +m^{IJ}(h^J)\right)\ , 	\\
0& = d\left(h^I-\frac12 db^I-\frac12\bar\star db^I\right)\wedge \bar\omega_I   \nonumber\\
&\qquad+\frac12 d\left((1+\bar\star)\left((h^I+m^{IJ}(h^J) -d a^I \right)\wedge (\omega_I-\bar\omega_I)\right)\ , \\
0& =d\bar\star\left(h^I+m^{IJ}(h^J)- d a^I\right) \ .
\end{align}
With this ansatz we find the action
\begin{align}
	S = &-\bar V\int_\Sigma \frac12 db^I\wedge \bar\star db^I + 2\eta_{IJ}h^I\wedge db^J  - \eta_{IJ}h^I\wedge  m^{JK}(h^K)	\nonumber\\
 &- \Omega_{IJ} \int_\Sigma \frac12 \left(d  a^I -h^I-m^{IK}(h^K)\right)\wedge \bar\star  \left(d  a^J -h^J-m^{JL}(h^L)\right)\ ,
\end{align}
where 
\begin{align}
	\Omega_{IJ} = \int_{K3} (\omega_I-\bar \omega_I)\wedge\bar\star (\omega_J-\bar \omega_J) \ .
\end{align}
In this was we find a sum over 22 copies of the  two-dimensional Sen action, generalised to include additional non-dynamical scalars $a^I$. 19 of these give self-dual fields and 3 anti-self-dual fields.     

\section{The M5-brane on a Riemann-Surface}\label{sec: RS}
 
Let us now consider reduction to four-dimensions. For a torus this case has been studied already in \cite{Sen:2019qit} (see also \cite{Aggarwal:2025fiq,Vanichchapongjaroen:2025psm}). However with the Hull formulation \cite{Hull:2023dgp} we can consider reduction on a genus $g$ Riemann-surface. This case includes the well-studied theories of class $\cal S$ \cite{Gaiotto:2009hg,Gaiotto:2009we}.

Following the discussion above we introduce two sets of harmonic 1-forms for the Riemann surface which can be chosen to be $i$-self-dual (recall that a bar does not denote complex conjugation):
\begin{align}
    \bar{\omega}_I&=i\bar{\star}\bar{\omega}_I\ ,\\
    \omega_I&=i\star\omega_I\ ,\\
    d\bar{\omega}_I&=d\bar{\star}\bar{\omega}_I=0\ ,\\
    d\omega_I&=d\star\omega_I=0 \ ,
\end{align}
where now $I=1,...,g$. It is helpful to split
\begin{align}
B = B_1+B_{0,2}	
\end{align}
where $B_1$ is expanded in terms of 1-forms on the Riemann surface
and $B_{0,2}$ be expanded in terms of 0-forms and 2-forms. 
Similarly we expand $H$ into two terms:
\begin{align}
	H = H_1+H_{0,2}\ .
\end{align}
Since the action is quadratic it also splits:
\begin{align}
	S = S_1+S_{0,2}\ ,
\end{align}
and it is simpler to look at $S_1$ and $S_{0,2}$ separately. 

For $S_1$ we  take
\begin{align}    
B_1&=(\bar{b}^I)^*\wedge\hspace{2pt}\bar{\omega}_I+(\bar{a}^I)^*\wedge (\omega_I-\bar\omega_I)+ c.c. \ ,\end{align}
and
\begin{align}
H_1&=(h^I)^*\wedge\bar{\omega}_I+\frac{1}{2}\left(1+\bar{\star}_{\bar{g}}\right)\left[(h^I+m(h^I))^*\wedge (\omega_I-\bar\omega_I)\right]+c.c.\ ,\\
    \cM(H_1)&=m(h^I)^*\wedge\bar{\omega}_I+\frac{1}{2}\left(1-\bar{\star}_{\bar{g}}\right)\left[(h^I+m(h^I))^*\wedge (\omega_I-\bar\omega_I)\right]+c.c.\ ,
\end{align}
so that
\begin{align}
H_1 + \cM(H_1) = 	(h^I+m(h^I))^*\wedge \omega_I + c.c.\ .
\end{align}
with the requirement that $h^I + m(h^I)=i\star (h^I + m(h^I))$.

Imposing that $H_1=\bar{\star}H_1$, and $M=-\bar{\star}M$, we have:
\begin{align}
h^I&=i\bar{\star}h^I\ ,\\
    m(h^I)&=-i\bar{\star}m(h^I)    \ ,
\end{align}
 With this ansatz we find the consistent solutions 
\begin{align}
0&=2im(h^I)-\bar{\star}db^I-idb^I\ , \\ 
0&= d(h^I-\frac{1}{2}i\bar{\star}db^I)\ ,\\[8pt]
    \bar{\star}da^I&=\bar{\star}(h^I+m(h^I))\ .
\end{align}
Substituting into the action we find
\begin{align}
S_1=-\int_{\Sigma} \frac{1}{2} d(\bar{b}^I)^* \wedge \bar{\star} d \bar{b}^J\Gamma_{IJ}&-2i(h^I)^*\wedge d \bar{b}^J\Gamma_{IJ}+i(h^I)^* \wedge m(h^J)\Gamma_{IJ}\nonumber\\[5pt]
+\frac{1}{2} d\bar({a}^I)^* \wedge \bar{\star} d \bar{a}^J\Omega_{IJ}&+\bar{\star}(h^I+m(h^I))^*\wedge d\bar{a}^J\Omega_{IJ}+(h^I)^*\wedge \bar{\star}m(h^J)\Omega_{IJ}+c.c.\label{rieact}
\end{align}
where
\begin{align}
 \Gamma_{IJ}&= \int_{\cal K}\bar{\omega}_I\wedge\bar{\star}(\bar{\omega}_J)^*\ ,\\ \
\Omega_{IJ} & = 	\int_{\cal K} d\chi_I\wedge\bar{\star}(d\chi_J)^*\ .
\end{align}
We find the equations of motion to be
\begin{align}
    d(h^I-\frac{1}{2}i\bar{\star}db^I)&=0\ ,\\[8pt]
    d\bar{\star}(h^I+m(h^I))&=d\bar{\star}da^I\ ,
\end{align}
and 
\begin{align}
&2im(h^I)-\bar{\star}db^I-idb^I\nonumber\\[4pt]
-&i(\Gamma^{-1}\Omega)_{IJ}m(\frac{1}{2}(1+i\bar{\star})da^J)-\frac{1}{2}(\Gamma^{-1}\Omega)_{IJ}\bar{\star}da^J-\frac{1}{2}i(\Gamma^{-1}\Omega)_{IJ}da^J\nonumber\\[8pt]
+&2i(\Gamma^{-1}\Omega)_{IJ}m(h^J)=0\ .\label{gen}
\end{align}
As before the equations of motion that follow from this action are more general than the consistent solution we found above. In particular the general solution of $a^I$ is 
\begin{equation}
    da^I=h^I+m(h^I)+i\bar{\star}d\bar{a}^I\ ,
\end{equation}
for an arbitrary  choice of $\bar{a}^I$. Nevertheless, as with the previous cases,  we can identify
\begin{align}
    h'^I&=h^I-\frac{1}{4}\Gamma^{-1}_{IK}\Omega_{KJ}(1+i\bar{\star})d\bar{a}^J\ ,\\[4pt]
    b'^I&=b^I-\frac{1}{2}\Gamma^{-1}_{IK}\Omega_{KJ}\bar{a}^J\ .
\end{align}
This identification absorbs $\bar{a}^I$ in the general solution, leaving us with something that is equivalent to the consistent solution:
\begin{align}
    d\left(h'^I-\frac{1}{2}(1+i\bar{\star})db'^I\right)&=0\ ,\\
    d\left(h'^I+m(h'^I)\right)&=0\ .
\end{align}

Next we turn to $S_{0,2}$ which is constructed from  0-forms and 2-forms on the Riemann Sirface. Note that there is a  single harmonic 0-form on the Riemann surface, namely the identity function $1$,  which is   closed and co-closed with respect to both metrics:
\begin{align}
d1 = d\bar \star 1 = d\star 1 =0 	\ .
\end{align} 
On the other hand there are two harmonic 2-forms $\star 1$ and $\bar \star 1$ associated to each of the two metrics. These are related by
\begin{align}
\star 1 = \sqrt\frac{\det \kappa}{\det \bar  \kappa}\	\bar \star 1\ ,
\end{align}
so they will in general have a different functional dependence on the coordinates of ${\cal K}$. 

This time we are lead to consider 
\begin{align}
B_{0,2} & = b_2 + b_0 \bar \star 1 + a_0 \star 1  \ .
\end{align} 
We find that a suitable expansions for $H_{0,2}$  and $M(H_{0,2})$ are
\begin{align}
H_{0,2} &= \frac12\left(\sqrt\frac{\det \kappa}{\det \bar  \kappa}\star + \bar \star \right)h_3\wedge \bar\star 1 + 	\frac12\left(\sqrt\frac{\det \kappa}{\det \bar  \kappa}\bar\star \star +1\right)h_3\ , \\
\cM(H_{0,2}) &= \frac12\left(\sqrt\frac{\det \kappa}{\det \bar  \kappa}\star - \bar \star \right)h_3\wedge \bar\star 1 -	\frac12\left(\sqrt\frac{\det \kappa}{\det \bar  \kappa}\bar \star \star - 1\right)h_3 \ .
\end{align}
where here $h_3$ is a 3-form on $\Sigma$. This leads to 
\begin{align}
F_{0,2}  &= H_{0,2} + \cM(H_{0,2})	\\
 & = h_3 + \star h_3\wedge \star 1\ ,
\end{align}
as desired and hence the equation of motion $dF_{0,2}=0$ is solved by   $dh_3=d\star h_3=0$. Examining the other six-dimensional equation (\ref{eqofm2}) of motion we also learn that
\begin{align}
h_3& =db_2 - \bar \star db_0	\\
\star h_3 & = da_0
\end{align}
These in turn imply that $d\bar\star db_0=0$ and $db_2 - \bar \star db_0 = \star da_0$. 

Note that on-shell we can write $db_0 = \bar\star dc_2 $ in which case we can absorb $c_2$ into $b'_2= b_2 - c_2$. 
We find a consistent KK truncation given by
\begin{align}
h_3 = db'_2 = \star da_0	\ .
\end{align}
Thus $b_0$ does not generate new degrees of freedom on shell. 
In particular we could have simply taken the more familiar ansatz:
\begin{align}
B_{0,2} & = b_2 +  a_0 \star 1  \ .
\end{align}   

Lastly let us compute $S_{0,2}$:
\begin{align}
S_{0,2}   = &-\bar V\int_{\Sigma}	\frac12db_2\wedge \bar\star db_2
+\frac12 db_0\wedge\bar\star db_0 + \bar\star h_3\wedge db_2 + h_3\wedge db_0\nonumber\\
&-V\int_{\Sigma}db_0\wedge\bar\star da_0 + \star h_3\wedge d
b_2 + \bar\star\star h_3\wedge db_0 +h_3\wedge da_0	\nonumber\\
&-\Omega\int_{\Sigma} \frac12 da_0\wedge \star da_0	+\bar\star\star h_3\wedge da_0 + \frac12 \star h_3\wedge\bar\star\star h_3\ ,\end{align}
where
\begin{align}
\bar V	& = \int_{\cal K}  \bar\star 1\ ,\\
V	&   = \int_{\cal K} \sqrt\frac{\det \kappa}{\det \bar  \kappa}\bar\star 1\ ,\\
\Omega  	&  = \int_{\cal K}  \frac{\det \kappa}{\det \bar  \kappa}  \bar \star 1  \ .
\end{align}
Assuming that  $\Omega\ne V^2/\bar V$ we find a more general set of equations 
\begin{align}
dh_3 &=d\bar \star db_0\ ,\\
d\bar\star\star h_3 & = d\bar\star da_0\ ,\\	
 Vd\star h_3 & = \bar V\left(d\bar \star db_2 - d\bar\star h_3\right)\ ,
\end{align}
and
\begin{align}\label{eqofm11}
 \bar V\bar\star \left(db_2 - \bar\star db_0-  h_3\right)+ V\star \left(db_2 -  \bar\star db_0 - \star da_0\right)+\Omega\star\bar\star\star\left( h_3-  \star  da_0\right)&=0 \ .	
\end{align}
 The consistent solutions correspond  to separately setting to zero  the coefficients of $\bar V, V$ and $\Omega$ in (\ref{eqofm11}). 
However these equations, which were derived from the reduced action,t   also admit inconsistent solutions, including solutions where $dh_3$ and $d\star h_3$ are non-vanishing. 

\section{Conclusion}\label{sec: Conclusion} 

In this paper we have studied how to compactify the  Sen action  \cite{Sen:2015nph,Sen:2019qit}  for self-dual fields, as extended to generic manifolds   by Hull \cite{Hull:2023dgp}. This theory is based on two independent metrics.  In order to consistently compactify the theory we saw that we needed to introduce a novel KK ansatz involving zero-modes obtained from the Laplacian's arising from  both metrics. This allowed us to construct consistent reductions. 

However it should be noted that the lower-dimensional theory admits additional solutions that do not lift to solutions of the higher-dimensional one. The reason for this is simple: The higher-dimensional equations of motion must be valid at each point in the compact manifold and this leads to a constraining set of equations that selects only certain lower-dimensional solutions. On the other hand if we substitute the KK reduction ansatz into the action and then integrate over the compact space then we only require that the equations of motion are satisfied 'on average' over the compact manifold which leads to less constraining equations and hence  a more general set of solutions. In this paper we focused on the six-dimensional case, associated to the M5-brane, and we hope to examine the reduction of ten-dimensional type IIB supergravity in future work. 

In our discussion we were led to observed that the Sen action admits an interesting generalization to include  an additional $2k$-form field which nevertheless does not lead to any new degrees of freedom, at least locally.  Nevertheless this field might be helpful in order to describe non-local operators such as Wilson lines and their higher dimensional generalizations and may lead to different dynamics in a non-trivial topological setting. We also hope to address these issues in future work.
  
\section*{Acknowledgements}

We would like to thank J.Lin,  I.  Bah and C. Hull for conversations. N.L. was supported in part by the STFC grant ST/X000753/1 and Y.Z. by the K-CSC studentship No.202308060068.


\begin{thebibliography}{0 }
 	
\bibitem{Siegel:1983es}
W.~Siegel,
Nucl. Phys. B \textbf{238} (1984), 307-316
doi:10.1016/0550-3213(84)90453-X
 	
\bibitem{Floreanini:1987as}
R.~Floreanini and R.~Jackiw,
Phys. Rev. Lett. \textbf{59} (1987), 1873
doi:10.1103/PhysRevLett.59.1873

\bibitem{Henneaux:1988gg}
M.~Henneaux and C.~Teitelboim,
Phys. Lett. B \textbf{206} (1988), 650-654
doi:10.1016/0370-2693(88)90712-5
	
\bibitem{Pasti:1996vs}
P.~Pasti, D.~P.~Sorokin and M.~Tonin,
Phys. Rev. D \textbf{55} (1997), 6292-6298
doi:10.1103/PhysRevD.55.6292
[arXiv:hep-th/9611100 [hep-th]].


\bibitem{Belov:2006jd}
D.~Belov and G.~W.~Moore,
[arXiv:hep-th/0605038 [hep-th]].

\bibitem{Mkrtchyan:2019opf}
K.~Mkrtchyan,
JHEP \textbf{12} (2019), 076
doi:10.1007/JHEP12(2019)076
[arXiv:1908.01789 [hep-th]].

\bibitem{Townsend:2019koy}
P.~K.~Townsend,
Phys. Rev. Lett. \textbf{124} (2020) no.10, 101604
doi:10.1103/PhysRevLett.124.101604
[arXiv:1912.04773 [hep-th]].

\bibitem{Sen:2015nph}
A.~Sen,
JHEP \textbf{07} (2016), 017
doi:10.1007/JHEP07(2016)017
[arXiv:1511.08220 [hep-th]].

\bibitem{Sen:2019qit}
A.~Sen,
J. Phys. A \textbf{53} (2020) no.8, 084002
doi:10.1088/1751-8121/ab5423
[arXiv:1903.12196 [hep-th]].
 	

\bibitem{Andriolo:2021gen}
E.~Andriolo, N.~Lambert, T.~Orchard and C.~Papageorgakis,
JHEP \textbf{04} (2022), 115
doi:10.1007/JHEP04(2022)115
[arXiv:2112.00040 [hep-th]].

\bibitem{Lambert:2023qgs}
N.~Lambert,
Phys. Lett. B \textbf{840} (2023), 137888
doi:10.1016/j.physletb.2023.137888
[arXiv:2302.10955 [hep-th]].

\bibitem{Hull:2025rxy}
C.~Hull and N.~Lambert,
[arXiv:2508.02865 [hep-th]].
 	
\bibitem{Hull:2023dgp}
C.~M.~Hull,
[arXiv:2307.04748 [hep-th]].


\bibitem{Hull:2025bqo}
C.~Hull and N.~Lambert,
[arXiv:2508.00199 [hep-th]].



\bibitem{Andriolo:2020ykk}
E.~Andriolo, N.~Lambert and C.~Papageorgakis,
JHEP \textbf{04} (2020), 200
doi:10.1007/JHEP04(2020)200
[arXiv:2003.10567 [hep-th]].


\bibitem{Aggarwal:2025fiq}
A.~Aggarwal, S.~Chakrabarti and M.~Raman,
[arXiv:2504.16673 [hep-th]].

\bibitem{Sen:2017szq}
A.~Sen,
JHEP \textbf{02} (2018), 155
doi:10.1007/JHEP02(2018)155
[arXiv:1711.08468 [hep-th]].

\bibitem{Hull:2025mtb}
C.~Hull,
[arXiv:2508.03902 [hep-th]].




\bibitem{Lin:2024eqq}
J.~Lin, T.~Skrzypek and K.~S.~Stelle,
JHEP \textbf{03} (2025), 200
doi:10.1007/JHEP03(2025)200
[arXiv:2412.00186 [hep-th]].

\bibitem{Lin:2025ucv}
J.~Lin, T.~Skrzypek and K.~S.~Stelle,
JHEP \textbf{12} (2025), 083
doi:10.1007/JHEP12(2025)083
[arXiv:2508.00987 [hep-th]].

\bibitem{Lin:2026rfc}
J.~Lin, K.~S.~Stelle and D.~Waldram,
[arXiv:2603.24534 [hep-th]].

\bibitem{Lu:2006dh}
H.~Lu, C.~N.~Pope and K.~S.~Stelle,
Nucl. Phys. B \textbf{782} (2007), 79-93
doi:10.1016/j.nuclphysb.2007.05.017
[arXiv:hep-th/0611299 [hep-th]].

\bibitem{Vanichchapongjaroen:2025psm}
P.~Vanichchapongjaroen,
JHEP \textbf{10} (2025), 223
doi:10.1007/JHEP10(2025)223
[arXiv:2506.07219 [hep-th]].


\bibitem{Hull:1994ys}
C.~M.~Hull and P.~K.~Townsend,
Nucl. Phys. B \textbf{438} (1995), 109-137
doi:10.1201/9781482268737-24
[arXiv:hep-th/9410167 [hep-th]].

\bibitem{Witten:1995ex}
E.~Witten,
Nucl. Phys. B \textbf{443} (1995), 85-126
doi:10.1201/9781482268737-32
[arXiv:hep-th/9503124 [hep-th]].

\bibitem{Cherkis:1997bx}
S.~A.~Cherkis and J.~H.~Schwarz,
Phys. Lett. B \textbf{403} (1997), 225-232
doi:10.1016/S0370-2693(97)00360-2
[arXiv:hep-th/9703062 [hep-th]].

\bibitem{Gaiotto:2009hg}
D.~Gaiotto, G.~W.~Moore and A.~Neitzke,
Adv. Math. \textbf{234} (2013), 239-403
doi:10.1016/j.aim.2012.09.027
[arXiv:0907.3987 [hep-th]].

\bibitem{Gaiotto:2009we}
D.~Gaiotto,
JHEP \textbf{08} (2012), 034
doi:10.1007/JHEP08(2012)034
[arXiv:0904.2715 [hep-th]].

\end{thebibliography}
 \end{document}